\documentclass{emulateapj}

\newcommand{\kms}{\ifmmode {\rm km\ s}^{-1} \else km s$^{-1}$\fi}
\newcommand{\Msun}{\ifmmode {\rm M}_{\odot} \else M$_{\odot}$\fi}
\newcommand{\Lsun}{\ifmmode {\rm L}_{\odot} \else L$_{\odot}$\fi}
\newcommand{\qo}{\ifmmode q_{\rm o} \else $q_{\rm o}$\fi}
\newcommand{\Ho}{\ifmmode H_{\rm o} \else $H_{\rm o}$\fi}
\newcommand{\ho}{\ifmmode h_{\rm o} \else $h_{\rm o}$\fi}

\newcommand{\vFWHM}{\ifmmode v_{\mbox{\tiny FWHM}} \else
                    $v_{\mbox{\tiny FWHM}}$\fi}
\newcommand{\CCF}{\ifmmode F_{\it CCF} \else $F_{\it CCF}$\fi}
\newcommand{\ACF}{\ifmmode F_{\it ACF} \else $F_{\it ACF}$\fi}
\newcommand{\Halpha}{\ifmmode {\rm H}\alpha \else H$\alpha$\fi}
\newcommand{\Hbeta}{\ifmmode {\rm H}\beta \else H$\beta$\fi}
\newcommand{\Hgamma}{\ifmmode {\rm H}\gamma \else H$\gamma$\fi}
\newcommand{\Hdelta}{\ifmmode {\rm H}\delta \else H$\delta$\fi}
\newcommand{\Lya}{\ifmmode {\rm Ly}\alpha \else Ly$\alpha$\fi}
\newcommand{\Lyb}{\ifmmode {\rm Ly}\beta \else Ly$\beta$\fi}
\newcommand{\HeI}{\ifmmode {\rm He}\,{\sc i}\,\lambda5876 \else 
	          He\,{\sc i}\,$\lambda5876$\fi}
\newcommand{\HeII}{\ifmmode {\rm He}\,{\sc ii}\,\lambda4686 \else 
	           He\,{\sc ii}\,$\lambda4686$\fi}

\newcommand{\ciii}{\ifmmode {\rm C}\,{\sc iii} \else C\,{\sc iii}\fi}

\newcommand{\oiii}{O\,{\sc iii}}

\shorttitle{Diverse BLR Velocity Fields}
\shortauthors{}

\accepted{2009 September 15}

\begin{document}

\title{Diverse Kinematic Signatures From Reverberation Mapping of the
Broad-Line Region in Active Galactic Nuclei}

\author{ K.~D.~Denney\altaffilmark{1}, B.~M.~Peterson\altaffilmark{1,2},
         R.~W.~Pogge\altaffilmark{1,2}, A.~Adair\altaffilmark{3},
         D.~W.~Atlee\altaffilmark{1}, K.~Au-Yong\altaffilmark{3},
         M.~C.~Bentz\altaffilmark{1,4}, J.~C.~Bird\altaffilmark{1},
         D.~J.~Brokofsky\altaffilmark{5,6}, E.~Chisholm\altaffilmark{3},
         M.~L.~Comins\altaffilmark{1,7}, M.~Dietrich\altaffilmark{1},
         V.~T.~Doroshenko\altaffilmark{8,9,10},
         J.~D.~Eastman\altaffilmark{1}, Y.~S.~Efimov\altaffilmark{9},
         S.~Ewald\altaffilmark{3}, S.~Ferbey\altaffilmark{3},
         C.~M.~Gaskell\altaffilmark{5,11},
         C.~H.~Hedrick\altaffilmark{5,7}, K.~Jackson\altaffilmark{3},
         S.~A.~Klimanov\altaffilmark{9,10},
         E.~S.~Klimek\altaffilmark{5,12}, A.~K.~Kruse\altaffilmark{5,13},
         A.~Lad\'{e}route\altaffilmark{3}, J.~B.~Lamb\altaffilmark{14},
         K.~Leighly\altaffilmark{15}, T.~Minezaki\altaffilmark{16},
         S.~V.~Nazarov\altaffilmark{9,10}, C.~A.~Onken\altaffilmark{17,18}
         E.~A.~Petersen\altaffilmark{5}, P.~Peterson\altaffilmark{19},
         S.~Poindexter\altaffilmark{1}, Y.~Sakata\altaffilmark{20}
         K.~J.~Schlesinger\altaffilmark{1},
         S.~G.~Sergeev\altaffilmark{9,10}, N.~Skolski\altaffilmark{3},
         L.~Stieglitz\altaffilmark{3}, J.~J.~Tobin\altaffilmark{14},
         C.~Unterborn\altaffilmark{1},
         M.~Vestergaard\altaffilmark{21,22},
         A.~E.~Watkins\altaffilmark{5}, L.~C.~Watson\altaffilmark{1},
         and Y.~Yoshii\altaffilmark{16} }

\altaffiltext{1}{Department of Astronomy, 
		The Ohio State University, 
		140 West 18th Avenue, 
		Columbus, OH 43210, USA; 
		denney, peterson,
                pogge@astronomy.ohio-state.edu}

\altaffiltext{2}{Center for Cosmology and AstroParticle Physics, 
                 The Ohio State University,
		 191 West Woodruff Avenue, 
		 Columbus, OH 43210, USA}

\altaffiltext{3}{Centre of the Universe,
                 Herzberg Institute of Astrophysics,
		 National Research Council of Canada,
		 5071 West Saanich Road,
		 Victoria, BC  V9E 2E7,
		 Canada}

\altaffiltext{4}{Present address: 
		 Department of Physics and Astronomy,
		 4129 Frederick Reines Hall,
		 University of California at Irvine,
		 Irvine, CA 92697-4575, USA;
		 mbentz@uci.edu}

\altaffiltext{5}{Department of Physics \& Astronomy, 
		 University of Nebraska, 
		 Lincoln, NE 68588-0111, USA. }

\altaffiltext{6}{Deceased, 2008 September 13}

\altaffiltext{7}{Current address: 
		 Astronomy and Astrophysics Department, 
		 Pennsylvania State University, 
		 525 Davey Laboratory, University Park, PA 16802, USA}

\altaffiltext{8}{Crimean Laboratory of the Sternberg Astronomical Institute, 
	         p/o Nauchny, 98409 Crimea, Ukraine;
		 vdorosh@sai.crimea.ua}

\altaffiltext{9}{Crimean Astrophysical Observatory,
		 p/o Nauchny, 98409 Crimea, Ukraine;
		 sergeev, efim@crao.crimea.ua,
		 sergdave2004@mail.ru,nazarastron2002@mail.ru}

\altaffiltext{10}{Isaak Newton Institute of Chile,
	          Crimean Branch, Ukraine}

\altaffiltext{11}{Current address: 
		 Astronomy Department, 
		 University of Texas, 
		 Austin, TX 78712-0259, USA;
                 gaskell@astro.as.utexas.edu}

\altaffiltext{12}{Current address: 
	          Astronomy Department, MSC 4500,
		  New Mexico State University, 
		  PO BOX 30001, La Cruces, NM 88003-8001, USA}

\altaffiltext{13}{Current address: 
                  Physics Department,
                  University of Wisconsin-Madison,
                  1150 University Avenue,
                  Madison, WI 53706-1390, USA}

\altaffiltext{14}{Department of Astronomy,
		 University of Michigan,
		 500 Church St., 
		 Ann Arbor, MI 48109-1040, USA}

\altaffiltext{15}{Homer L. Dodge Department of Physics and Astronomy,
	    	  The University of Oklahoma,
  		  440 W. Brooks St.,
  		  Norman, OK 73019, USA}
	
\altaffiltext{16}{Institute of Astronomy, 
		 School of Science, 
		 University of Tokyo,
	 	 2-21-1 Osawa, Mitaka, 
		 Tokyo 181-0015, Japan;
		 minezaki, yoshii@ioa.s.u-tokyo.ac.jp}

\altaffiltext{17}{Plaskett Fellow; Dominion Astrophysical Observatory,
                  Herzberg Institute of Astrophysics, 
		  National Research Council of Canada,
		  5071 West Saanich Road, Victoria, BC V9E 2E7, 
		  Canada}

\altaffiltext{18}{Current address:
                 Mount Stromlo Observatory,
		 Research School of Astronomy \& Astrophysics,
		 The Australian National University,
		 Cotter Road,
		 Weston Creek, ACT 2611,
		 Australia; 
		 onken@mso.anu.edu.au}

\altaffiltext{19}{Ohio University,
		  Department of Physics and Astronomy,
		  Athens, OH 45701-2979, USA}

\altaffiltext{20}{Department of Astronomy, 
		  School of Science, 
		  University of Tokyo,
  		  7-3-1 Hongo, Bunkyo-ku, 
		  Tokyo 113-0013, Japan}

\altaffiltext{21}{Steward Observatory, 
		The University of Arizona, 
		933 North Cherry Avenue, 
         	Tucson, AZ 85721, USA}

\altaffiltext{22}{DARK Cosmology Centre,
                 Niels Bohr Institute,
                 Copenhagan University}

\begin{abstract}
A detailed analysis of the data from a high sampling rate, multi-month
reverberation mapping campaign, undertaken primarily at MDM Observatory
with supporting observations from telescopes around the world, reveals
that the \Hbeta\ emission region within the broad line regions (BLRs) of
several nearby AGNs exhibit a variety of kinematic behaviors.  While the
primary goal of this campaign was to obtain either new or improved
\Hbeta\ reverberation lag measurements for several relatively low
luminosity AGNs, we were also able to unambiguously reconstruct
velocity-resolved reverberation signals from a subset of our targets.
Through high cadence spectroscopic monitoring of the optical continuum
and broad \Hbeta\ emission line variations observed in the nuclear
regions of NGC\,3227, NGC\,3516, and NGC\,5548, we clearly see evidence
for outflowing, infalling, and virialized BLR gas motions, respectively.
\end{abstract}



\section{INTRODUCTION}

Reverberation mapping \citep{Blandford82,Peterson93} is a technique
applied to spectroscopic observations of type 1 AGNs to infer properties
of the broad line-emitting region (BLR) through characterizations of
time delays between continuum and broad emission-line flux variations.
This method has become extremely useful and quite successful in its
current application of directly measuring BLR radii and black hole
masses ($M_{\rm BH}$).  However, the fundamental objective of
reverberation mapping, as its name implies, is to reconstruct or map the
emissivity and velocity distribution of the BLR line-emitting gas as a
function of position as it `reverberates' in response to the flux
variations of the ionizing continuum.  Because the position of the gas
can be inferred through the emission-line time delay, $\tau$ ($R_{\rm
BLR} = c\tau$), the resulting reconstruction is called a velocity--delay
map \citep[see][]{Horne04} and is the best means, with current
technology, to obtain direct knowledge about the geometry and kinematics
of the BLR.

Past attempts at producing velocity--delay maps \citep[e.g.,][]{Done96,
Ulrich96, Kollatschny03} have not yielded completely satisfactory
results, primarily on account of limitations in temporal sampling, with
inadequate time resolution or campaign duration or both.  Even given
these limitations, previous studies have successfully measured time
delays and black hole masses in over 40 type 1 AGNs \citep[see
e.g.,][]{Peterson04, Bentz09c}.  In addition, basic investigations into
time delay differences between multiple emission lines have shown that
the BLR is virialized across broad line-emitting regions of different
species \citep[e.g.,][and references therein]{Peterson04}.  Other
studies of the velocity dependence of the lag across a single emission
line region have shown suggestive evidence that the BLR commonly
contains a radial inflow component in addition to circular motions
\citep[e.g.,][]{Gaskell88, Crenshaw90, Koratkar91c, Done96}.

The experiences from earlier reverberation programs have led to more recent
campaigns that address the main observational obstacles encountered in
the past --- relative sampling rate, campaign duration, and data
quality.  Consequently, there has been consistent success in measuring
BLR radii in new targets and remeasuring radii that now supersede
previous, lower-precision or ambiguous measurements due to inadequate
time-sampling \citep[e.g.,][hereafter D09a]{Denney06, Bentz06b, Grier08,
Denney09b}.  Furthermore, these campaigns enable statistically
significant detections of reverberation signals at higher velocity
resolutions than have previously been found due mainly to the data
restrictions discussed above.  For example, \citet{Bentz08} revealed a
clear signal of radial inflow within the \Hbeta\ emission region of the
BLR in Arp~151, and \citet{Bentz09c} show further evidence for distinct
kinematic behavior in at least one other AGN from the same program (the
Lick AGN Monitoring Project).  Here we report on results of our own
recent high time-resolution monitoring program undertaken at five
observatories.  The first results of this program have already been
published (D09a), and additional results are in preparation
(K. D. Denney et al. 2009, in preparation, hereafter D09b).  In this
Letter, we present the most clear velocity-resolved reverberation
signals recovered from this campaign.  Our results highlight three
distinctly different BLR kinematic signatures --- inflow, outflow, and
virialized motions --- for three separate targets --- NGC\,3227,
NGC\,3516, and NGC\,5548, respectively.  These results reveal kinematic
diversity in reverberation signals and underscore the importance of
higher time resolution spectral monitoring.  This represents an
important step between measuring average BLR response times and black
hole masses to realizing the full potential of this technique through
the recovery of a velocity--delay map.

\section{Observations and Data Analysis}

\begin{figure}
\figurenum{1}
\epsscale{1.15}
\plotone{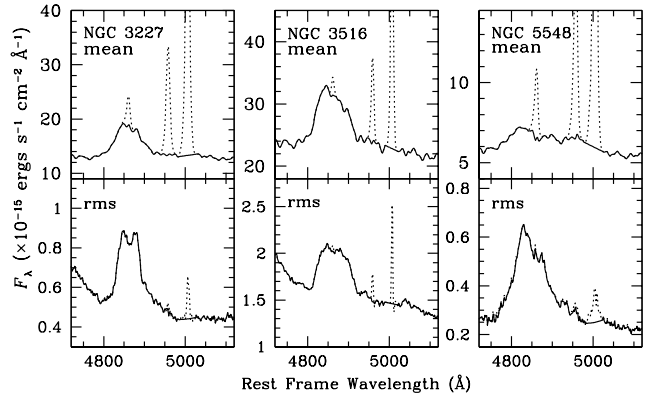}

\caption{Mean and rms spectra of NGC\,3227 (left), NGC\,3516 (middle),
and NGC\,5548 (right) from MDM observations.  The solid line shows the
mean and rms spectrum formed after removal of the [O\,{\sc
iii}]\,$\lambda\lambda 4959, 5007$ narrow emission lines, and the dotted
lines represent the spectra prior to this subtraction (note the small
residuals in the rms spectra).}

\label{fig:meanrms}
\end{figure}

Spectra of the nuclear regions of NGC\,3227, NGC\,3516, and NGC\,5548
were obtained from a combination of the 1.3 m telescope at MDM
Observatory, the 2.6 m Shajn telescope of the Crimean Astrophysical
Observatory (CrAO), and the Plaskett 1.8 m telescope at Dominion
Astrophysical Observatory (DAO).  Spectroscopic observations targeted
the H$\beta\,\lambda 4861$ and [O\,{\sc iii}]\,$\lambda\lambda 4959,
5007$ emission line region of the optical spectrum.  The top panels of
Figure \ref{fig:meanrms} show the mean spectrum of each of the three
targets based on the MDM observations, and the bottom panels show only
the variable emission in the form of an rms spectrum for each object,
respectively.  Emission line light curves were made from the integrated
\Hbeta\ flux measured above a linearly interpolated continuum fit to the
MDM, CrAO, and DAO spectra.

\begin{figure}
\figurenum{2}
\epsscale{1.15}
\plotone{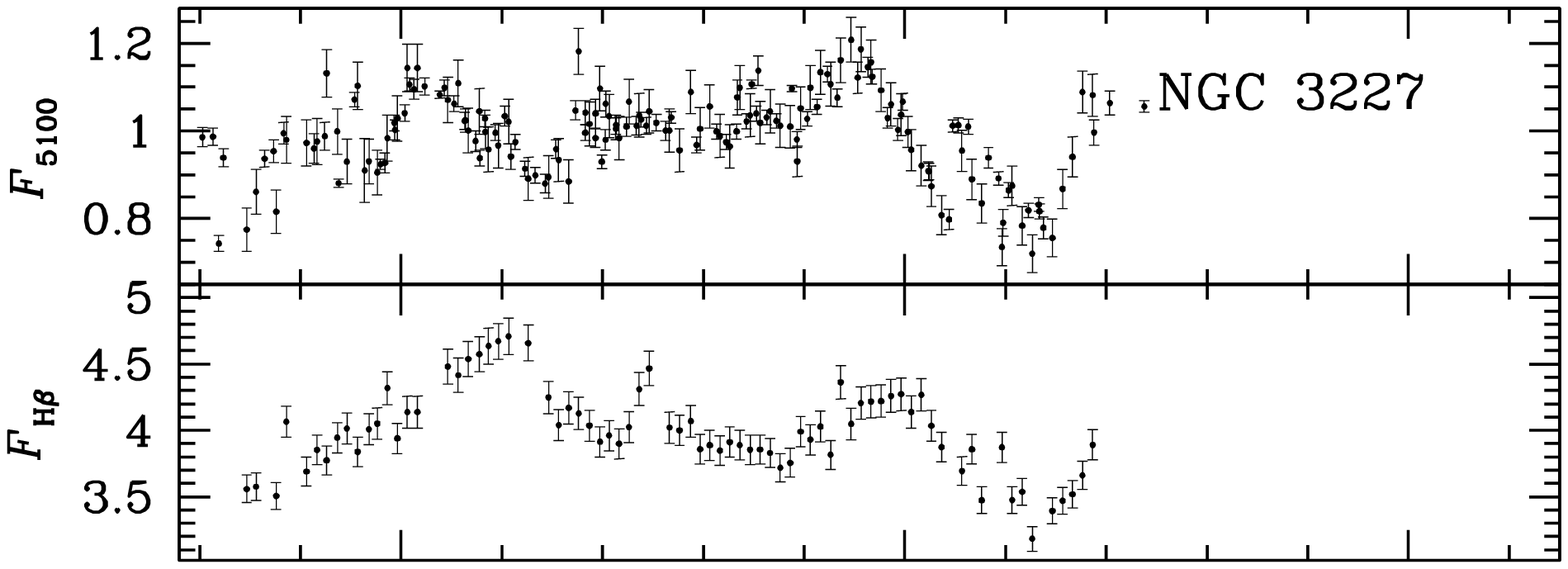}\\
\plotone{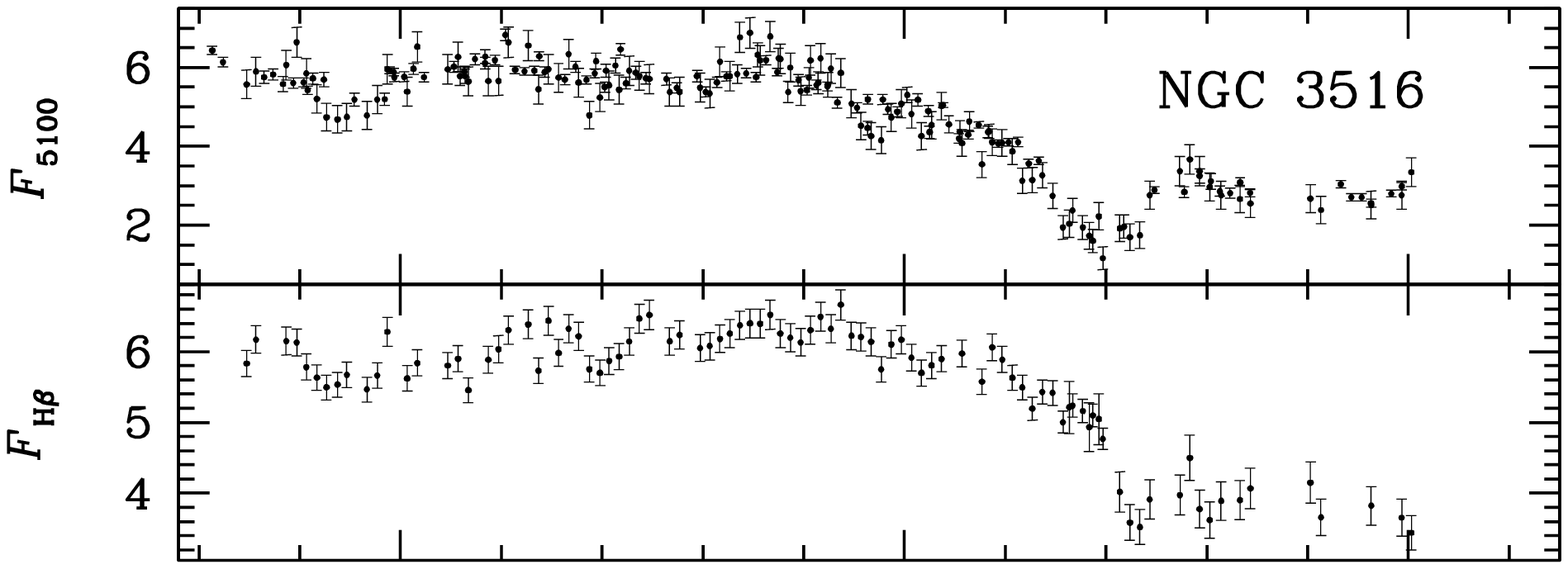}\\
\plotone{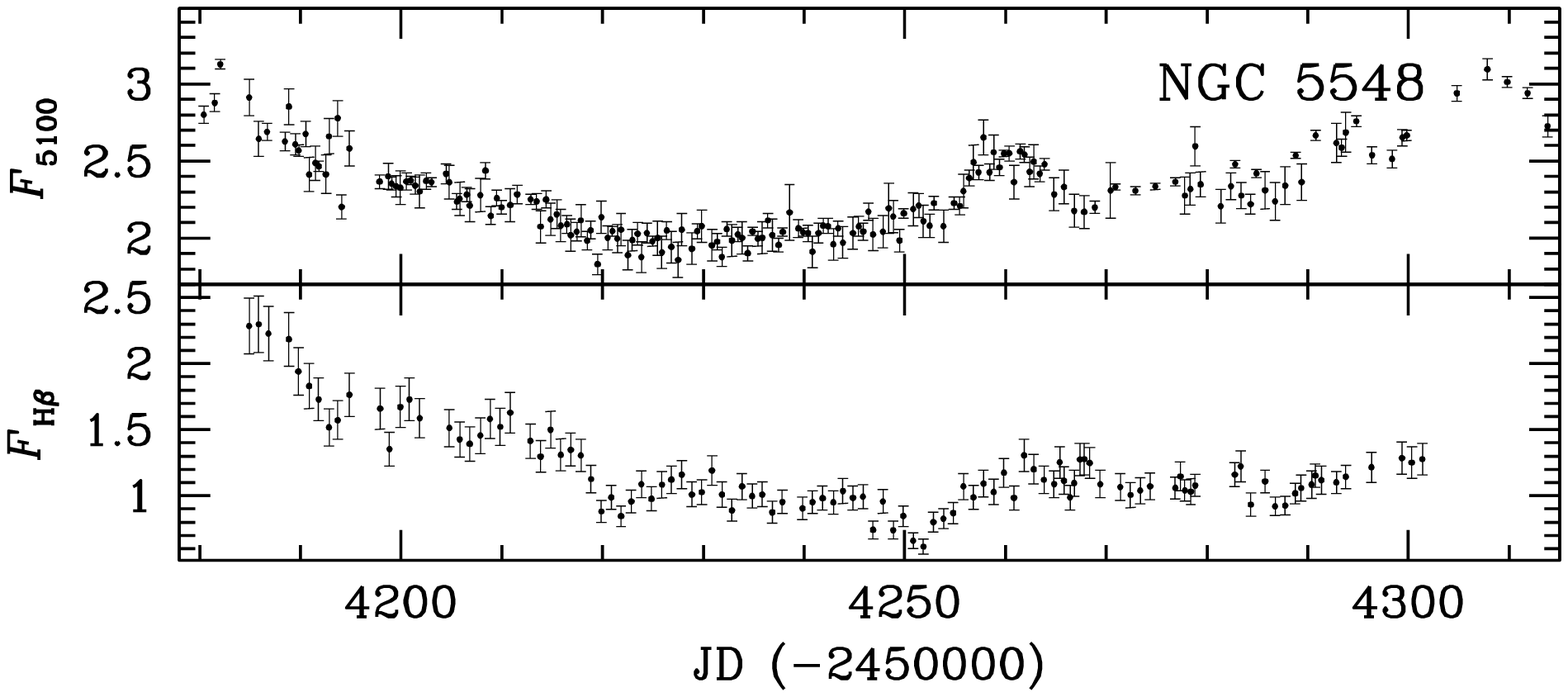}

\caption{Final light curves, after merging all data sets, of the
5100\AA\ continuum (top panels) and broad \Hbeta\ emission line flux
(bottom panels) in units of $10^{-15}$ erg s$^{-1}$ cm$^{-2}$ \AA$^{-1}$
and $10^{-13}$ erg s$^{-1}$ cm$^{-2}$, respectively, for NGC\,3227 (top),
NGC\,3516(middle), and NGC\,5548 (bottom).  Except for the continuum flux
scale of NGC\,3227, which is arbitrary because a linear fit to long-term
secular trends has been divided out, the flux scales reflect all
relative and absolute flux calibrations described by D09b.}

\label{fig:lc4cc}
\end{figure}

In addition to spectral observations, we obtained supplemental $V$-band
photometry from the 2.0-m Multicolor Active Galactic NUclei Monitoring
(MAGNUM) telescope at the Haleakala Observatories in Hawaii
\citep{Yoshii02, Yoshii03}, the 70-cm telescope of the CrAO, and the
0.4-m telescope of the University of Nebraska (UNebr.).  Continuum light
curves were created with observations from each $V$-band photometric
data set and the average continuum flux density near rest frame
$\sim$5100~\AA\ in each spectrum of the spectroscopic data sets.  The
reader is referred to D09a and D09b for details describing campaign
observing setups and data reduction, flux calibration of the spectra,
and intercalibration of the data sets to form the single set of optical
continuum and \Hbeta\ emission line light curves shown for each object
in Figure \ref{fig:lc4cc}.

\begin{deluxetable*}{lccccccc}
\tablecolumns{8}
\tablewidth{400pt}
\tablecaption{Light Curve Statistics and Mean \Hbeta\ Lags\label{tab:lcstats}
}
\tablehead{
\colhead{ } &
\colhead{Time} &
\colhead{ } &
\colhead{$T_{\rm median}$} &
\colhead{ } &
\colhead{Mean} &
\colhead{ } &
\colhead{ }\\
\colhead{Object} &
\colhead{Series} &
\colhead{$N$} &
\colhead{(days)} &
\colhead{Frac. Err} & 
\colhead{$F_{\rm var}$} &
\colhead{$R{\rm max}$} &
\colhead{$\tau_{\rm cent}$ (days)}\\
\colhead{(1)} &
\colhead{(2)} &
\colhead{(3)} &
\colhead{(4)} &
\colhead{(5)} &
\colhead{(6)} &
\colhead{(7)} &
\colhead{(8)}
}

\startdata

NGC\,3227 & $5100$ \AA & $171$ & $0.45$ & $0.03$ & $0.10$ & $1.9 \pm
0.1$ & \nodata \\ 
 & H$\beta$ & $75$ & $1.00$ & $0.03$ & $0.08$ & $1.5 \pm 0.1$ &
 $3.8 \pm 0.8$ \\
NGC\,3516 & $5100$ \AA & $198$ & $0.54$ & $0.06$ & $0.28$ & $5.9 \pm
1.5$ & \nodata \\ 
 & H$\beta$ & $93$ & $1.00$ & $0.04$ & $0.15$ & $1.9 \pm 0.2$ &
 $11.7^{+1.0}_{-1.5}$ \\
NGC\,5548 & $5100$ \AA & $182$ & $0.56$ & $0.03$ & $0.11$ & $1.7 \pm
0.1$ & \nodata \\ 
 & H$\beta$ & $108$ & $1.00$ & $0.09$ & $0.26$ & $3.7 \pm 0.5$ &
 $12.4^{+2.7}_{-3.9}$ \\

\enddata
\end{deluxetable*}

Table \ref{tab:lcstats} displays basic statistical parameters describing
the final light curves shown in Figure \ref{fig:lc4cc}.  Column 1
gives the object, and Column 2 lists the spectral feature represented
by each light curve.  The number of data points in each light curve is
shown in Column 3, with the median sampling interval between these
data points given in Column 4.  Column 5 shows the mean fractional
error in the fluxes of each time series.  Column 6 gives the excess
variance, calculated as

\begin{equation}
F_{\rm var} = \frac{\sqrt{\sigma^2 - \delta^2}}{\langle f \rangle}
\end{equation}

\noindent where $\sigma^2$ is the variance of the observed fluxes,
$\delta^2$ is their mean square uncertainty, and $\langle f \rangle$ is
the mean of the observed fluxes \citep{Rodriguezpascual97}.  Column 7
is the ratio of the maximum to minimum flux in the light curves, and
Column 8 gives the adopted mean \Hbeta\ time lag and uncertainties
determined through the primary time series analysis which utilized the
full \Hbeta\ line profile and is described in detail for each object by
D09b.

\section{Velocity-Resolved Time Series Analysis}
\label{S:velresInvest}

The lag measurements between the continuum and \Hbeta\ emission listed
in Table \ref{tab:lcstats} represent the average time delay across the
BLR, measured from the centroid of the cross correlation function
\citep[see][and references therein]{Peterson98} produced during the time
series analysis of the continuum and full \Hbeta\ line profile light
curves shown in Figure \ref{fig:lc4cc} (see D09a for details).  Here, we
focus on the velocity-resolved time series analysis we performed on each
target to investigate the potential for recovering velocity-dependent
time delays across the \Hbeta\ emission line in order to infer the
kinematic structure of the line emitting gas.

\begin{figure}
\figurenum{3}
\epsscale{1.15}
\plotone{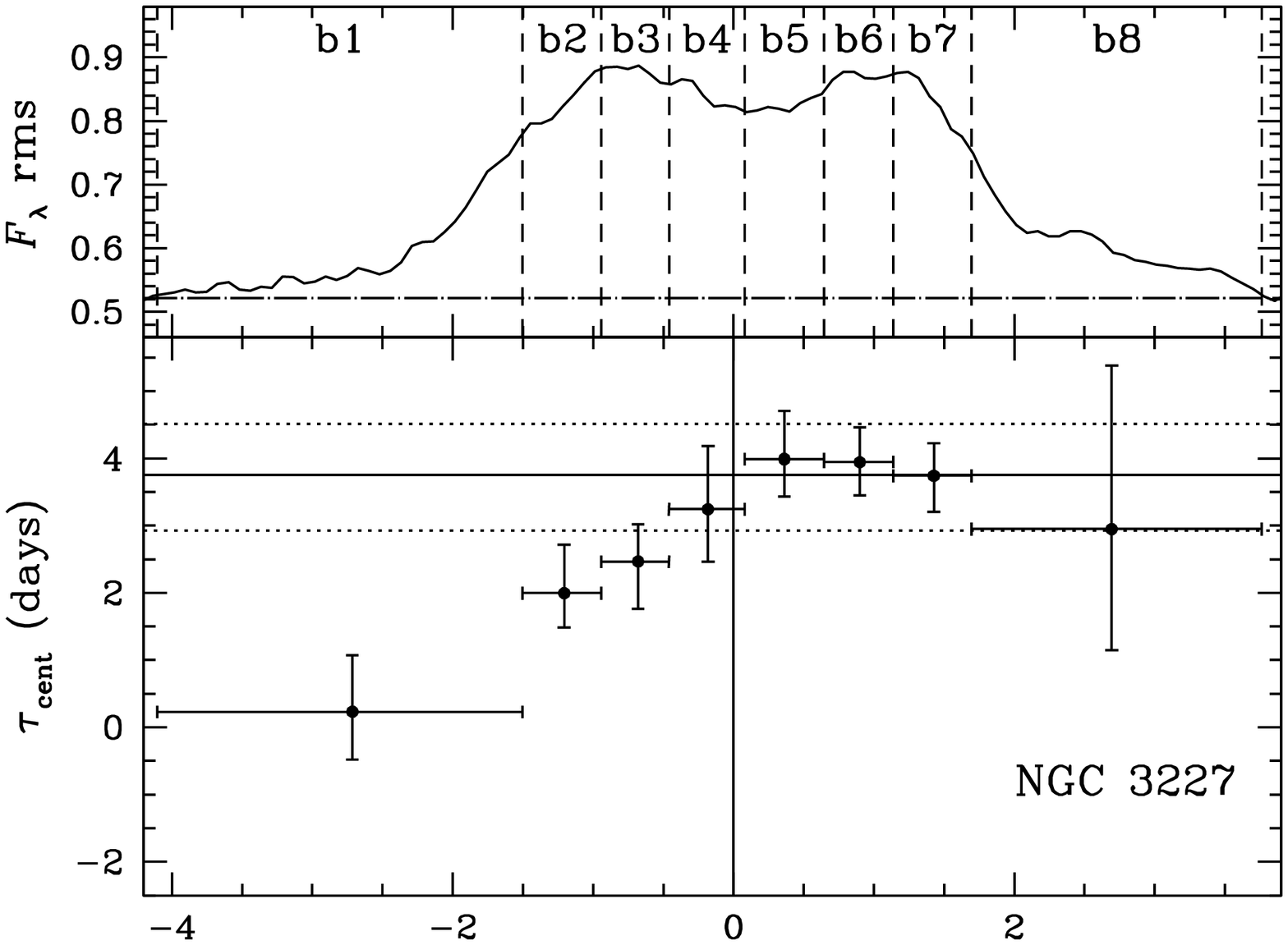}\\
\plotone{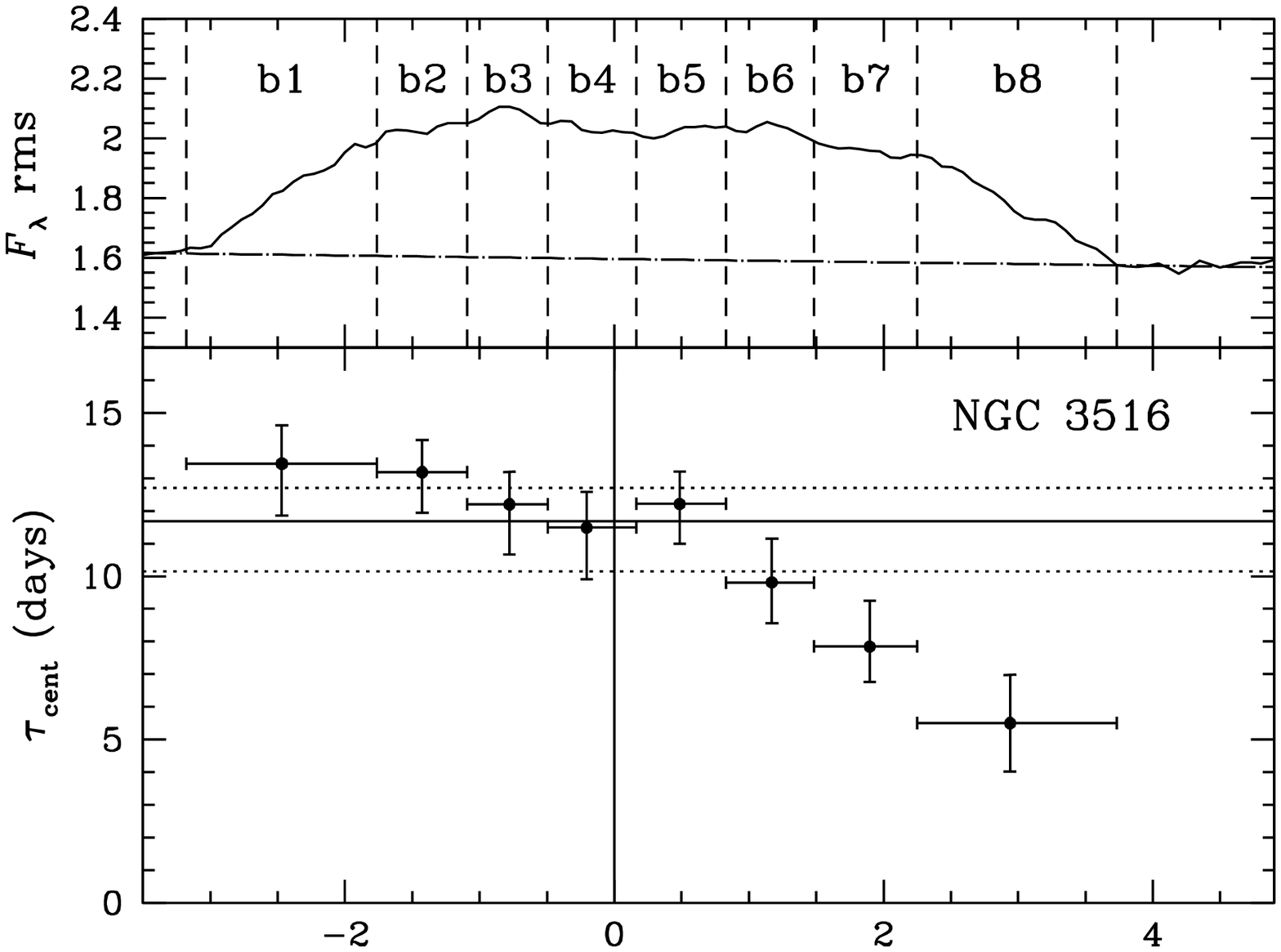}\\
\plotone{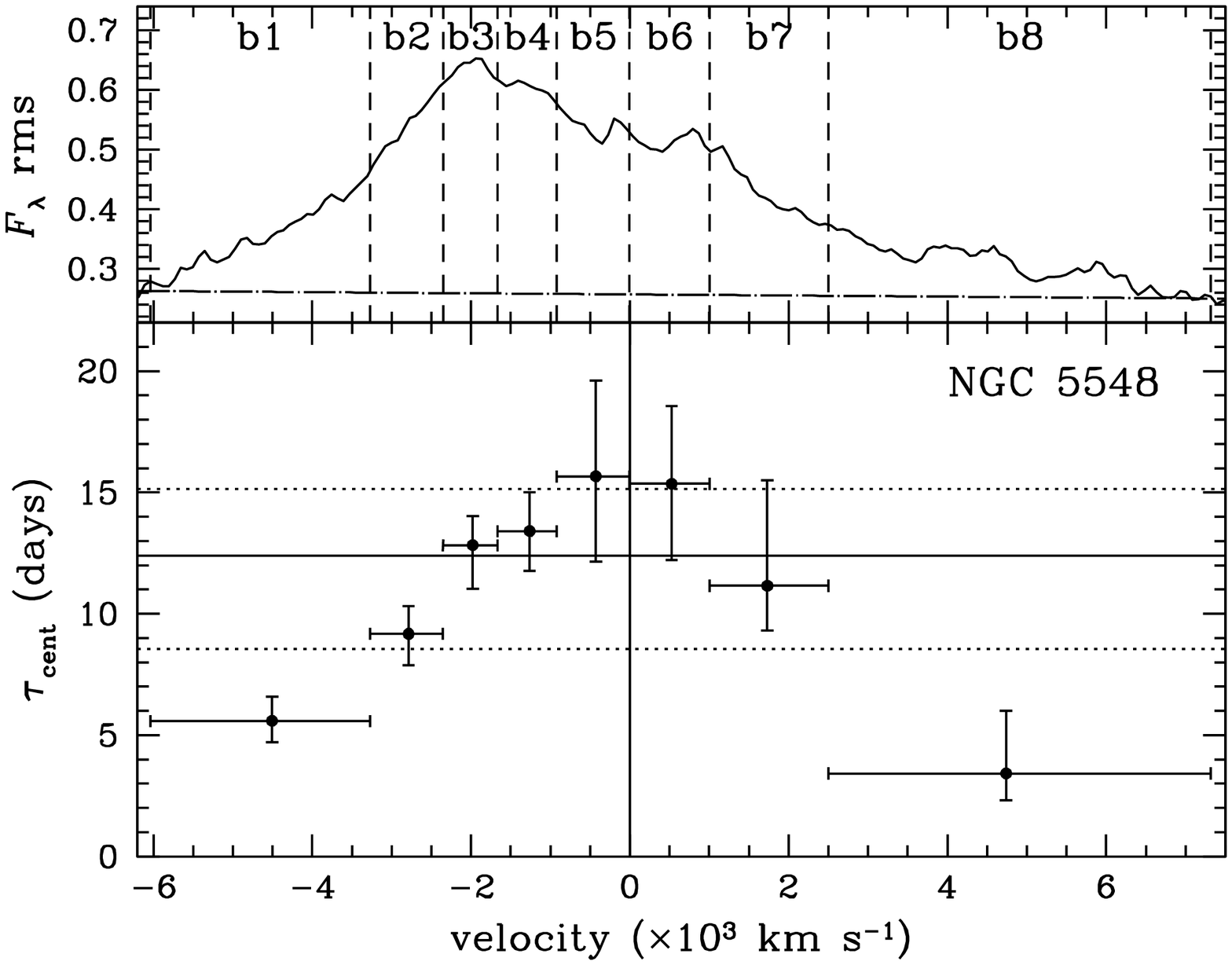}

\caption{Division of the \Hbeta\ rms spectral profile into equal-flux
bins (top panels; vertical dashed lines), and corresponding
velocity-resolved time-delay measurements (bottom panels) for NGC\,3227
(top), NGC\,3516 (middle), and NGC\,5548 (bottom), where the delays are
plotted at the flux centroid of each velocity bin.  Error bars on the lag
measurements in the velocity direction (bottom panels) reflect the bin
size, with each bin labeled by number in the top panels and negative
velocities referring to blueshifts from the line center and positive
velocities redshifts.  Error bars on the lag measurements are determined
similarly to those for the mean BLR lag (\citealt{Peterson98} with
modifications described by \citealt{Peterson04}).  The horizontal solid
and dotted lines in the bottom panel show the mean BLR lag and
associated errors, as listed in Table 1, while the horizontal
dotted-dashed line in the top panel represents the linearly-fit
continuum level.  Flux units are the same as in Fig. \ref{fig:meanrms}.}

\label{fig:reslags}
\end{figure}

We divided the \Hbeta\ emission line into eight velocity-space bins,
whose boundaries were determined by the division of the rms spectrum of
each object into eight bins of equal flux, as depicted in the top panels
of Figure \ref{fig:reslags}.  Compared to the line boundaries used for
the full profile analysis leading to the light curves shown in Figure
\ref{fig:lc4cc}, those used for this analysis were slightly narrowed in
the cases of NGC\,3516 and NGC\,3227 in order to include only the most
variable portions of the line profile, and boundaries were broadened for
NGC\,5548 because the rms spectrum shows variability in the red \Hbeta\
wing that extends beneath the [\oiii]\,$\lambda 4959$ narrow emission
line.

Light curves were created from measurements of the integrated \Hbeta\
flux in each bin and then cross correlated with the continuum light
curves shown in Figure \ref{fig:lc4cc}.  The bottom panels of Figure
\ref{fig:reslags} show the lag measurements for each of these bins.
Error bars in the velocity direction represent the bin width.  The
evidence for a velocity-stratified BLR response to continuum variations
is clear in all three cases.  Interestingly, each case demonstrates a
different kinematic signature: (1) outflow is indicated in NGC\,3227,
given the generally longer lags from red-shifted BLR gas compared to the
gas on the blue-shifted side of the line, (2) NGC\,3516 shows the
opposite signature, with the blue side of the line lagging the red side
--- an indication that there is an infall component to the gas in this
region, similar to that observed in Arp~151 by \citet{Bentz08}, and (3)
NGC\,5548 shows no radial gas motions, with relatively symmetric lags
measurements extending to equally large velocities on both the red and
blue sides of line --- a clear indication of virialized gas motions,
with the high-velocity line wings arising in gas closest to the central
source (see \citealt{Robinson90, Welsh&Horne91, Perez92, Bentz09c} for
examples of how velocity resolved responses can be related to different
BLR geometries; see \citet{Peterson01} for a related tutorial).

\section{Discussion}

The velocity-resolved reverberation signals exhibited by the
\Hbeta-emitting gas in the BLR of these three sources show striking
differences in kinematic behavior.  This is of particular interest
because black hole masses derived from reverberation mapping are only
valid under the assumption of gravitational domination of the BLR gas
dynamics by the black hole.  The influence of gravity on the BLR in
NGC\,3516 and NGC\,5548 is evidenced by the signatures of an infalling
gas component in the former and a lack of significant radial motions of
any kind in the latter (see Figure \ref{fig:reslags}).  The apparent
evidence for outflow in the case of NGC\,3227 is a first for
reverberation mapping.  In some sense, this should be no surprise given
the overwhelming evidence for large-scale mass loss from the inner
regions of AGNs \citep*{Crenshaw03}.  On the other hand, this does call
into question the assumptions that allow us to estimate the mass of the
central object, though we do hasten to point out that an outflow at
escape velocity would still allow us to measure the black hole mass with
the levels of accuracy currently claimed.  Furthermore, the black hole
mass we calculate for NGC\,3227 of $M_{\rm BH} = (7.6^{+1.6}_{-1.7})
\times 10^{6}M_{\odot}$, based on the lag in Table 1 and the line
dispersion of the broad \Hbeta\ line measured from the rms spectrum (see
D09b), is consistent within the statistical and systematic uncertainties
in this method to independent measurements using galactic stellar
\citep{Davies06} and gas \citep{Hicks08} dynamics.  We also note that
while the \Hbeta\ emission line is blueward asymmetric, the \Hbeta\
profiles of NGC\,3516 and NGC\,5548 are even more so, so we ascribe
little importance to this.  It strikes us as likely that we are
observing a complex system such as a two-component BLR, in which there
is an outflowing wind, in addition to a virialized, disk component
\citep{Murray&Chiang97, Eracleous&Halpern03}.  In this case, the mean
lag that we measure, which is tracing the position of the majority of
the line-emitting gas, arises from the disk component and is therefore
under the gravitational influence of the black hole.  Meanwhile, the
high-velocity, blueshifted emission with very short lags arises from a
wind at the inner BLR.  Additional support for this scenario comes from
the double-peaked profile shape of the rms spectrum, suggesting a
disk-like origin \citep[e.g.,][]{Eracleous&Halpern94}.  A further test
for the virial nature of the BLR in NGC\,3227 is to search for
reverberation signals from multiple emission lines in this object.  Data
from this same campaign suggest that the \HeII\ emission-line flux also
varied significantly over the course of the campaign, and future work is
planned to search for a reverberation signal from this variable line
emission.

The observations of virial motions in NGC\,5548 and the outflow in
NGC\,3227, in particular, are of further interest in the context of
comparisons with previous velocity-resolved studies of a similar nature
\citep[e.g.,][]{Sergeev99, Doroshenko08}.  These are largely focused on
NGC\,5548 (the object for which we have the most reverberation mapping
data), and a comprehensive summary of many of these past studies is
given by \citet{Gaskell&Goosmann08}, who discuss the support for
infalling BLR gas as implied by these various results, particularly for
low-ionization lines like \Hbeta.  This is in contrast to what we
clearly see in NGC\,5548 and NGC\,3227.  It is entirely possible that
the velocity fields in these regions could change as a function of
accretion rate, luminosity state, or over dynamical timescales.
Interpretation of these new results is problematic and probably will
remain so with only a handful of examples at single epochs.  At this
point, generalizing from these few sources is premature.

\section{Summary}

In this work, we have presented three clear cases of differing velocity
signatures of \Hbeta-emitting gas from the BLR of three nearby AGNs,
demonstrating the diversity and probable complexity of the kinematics in
this region.  Our ability with this work and that of \citet[][see also
\citealt{Bentz08}]{Bentz09c} to recover statistically significant
velocity-resolved reverberation responses for multiple objects is
principally due to successful completion of campaigns in which high
quality, homogeneous observations were obtained over long durations
(i.e., multiples of the reverberation lags) with a sampling rate that
was high compared to the relevant timescales being investigated.
Velocity-resolved reverberation mapping studies such as described here
are the next step toward producing a velocity--delay map, which will
reconstruct the two-dimensional kinematic structure of the BLR.  This,
in turn, will allow further insight into the geometry and dynamics of
the BLR, potentially leading to estimates of its inclination and
ultimately reducing the systematic uncertainties in $M_{\rm BH}$
determinations.  In particular, placing direct observational constraints
on the value of the reverberation mapping mass scale factor, $f$
\citep[see][]{Onken04}, even in individual objects, could reduce the
scatter in the $M_{\rm BH}$--$\sigma_{\star}$ relation for AGNs
\citep{Gebhardt00b, Ferrarese01, Onken04, Nelson04}.  Despite the
diversity in BLR kinematic signatures seen here, we do not see any
systematic trends in the location of these objects on the $M_{\rm
BH}$--$\sigma_{\star}$ relation, suggesting that any effects the
different velocity fields of these objects have on their black hole mass
estimates are within the current scatter in this relationship.  More
detailed information about the velocity fields of these objects is
needed to say anything further on the effect these kinematic differences
have on their black hole mass measurements.  Future work will focus on
revealing more structure in BLR velocity fields as we work to create
velocity--delay maps of the \Hbeta\ emission in NGC\,3227, NGC\,3516,
and NGC\,5548 from the data presented here.

\acknowledgements We acknowledge support for this work by the National
Science Foundation though grant AST-0604066 to The Ohio State
University.  CMG is grateful for support by the National Science
Foundation through grants AST 03-07912 and AST 08-03883.  MV
acknowledges financial support from HST grants HST-GO-10417,
HST-AR-10691, and HST-GO-10833 awarded by the Space Telescope Science
Institute, which is operated by the Association of Universities for
Research in Astronomy, Inc., for NASA, under contract NAS5-26555.  VTD
acknowledges the support of the Russian Foundation for Basic Research
(project no. 06-02-16843) to the Crimean Laboratory of the Sternberg
Astronomical Institute.  SGS acknowledges support through Grant No. 5-20
of the "Cosmomicrophysics" program of the National Academy of Sciences
of Ukraine to CrAO.  The CrAO CCD cameras have been purchased through
the US Civilian Research and Development Foundation for the Independent
States of the Former Soviet Union (CRDF) awards UP1-2116 and
UP1-2549-CR-03.  This research has made use of the NASA/IPAC
Extragalactic Database (NED) which is operated by the Jet Propulsion
Laboratory, California Institute of Technology, under contract with the
National Aeronautics and Space Administration and is based on
observations obtained at the Dominion Astrophysical Observatory,
Herzberg Istitute of Astrophysics, National Research Council of Canada.







\end{document}